\documentclass[aps,prl,reprint,twocolumn,superscriptaddress,floatfix]{revtex4-2}
\usepackage{amsmath}
\usepackage{amssymb}
\usepackage[utf8]{inputenc}
\usepackage{makeidx}
\usepackage{graphicx}
\usepackage[caption=false]{subfig}
\usepackage{hyperref} 
\usepackage{siunitx}
\usepackage{xcolor}
\usepackage{braket}
\usepackage{booktabs}
\usepackage{makecell}
\usepackage[normalem]{ulem}
\usepackage{newfloat}
\usepackage{dcolumn}

\makeatletter
\newcommand*{\balancecolsandclearpage}{%
  \close@column@grid
  \clearpage
  \twocolumngrid
} \makeatother

\pdfoutput=1

\begin{document}

\title{Spin-orbit coupling controlled 2D magnetism in chromium trihalides}

\author{Inhee Lee}\email{lee.2338@osu.edu}
\affiliation{Department of Physics, The Ohio State University,
Columbus, OH 43210, USA}

\author{Jiefu Cen}
\affiliation{Department of Physics, University of Toronto,
Toronto, Ontario, Canada M5S 1A7}

\author{Oleksandr Molchanov}
\affiliation{Department of Physics, The Ohio State University,
Columbus, OH 43210, USA}

\author{Shi Feng}
\affiliation{Department of Physics, The Ohio State University,
Columbus, OH 43210, USA}

\author{Warren L. Huey}
\affiliation{Department of Chemistry and Biochemistry, The Ohio
State University, Columbus, OH 43210, USA}

\author{Johan van Tol}
\affiliation{National High Magnetic Field Laboratory, Florida
State University, Tallahassee, FL 32310, USA}

\author{Joshua E. Goldberger}
\affiliation{Department of Chemistry and Biochemistry, The Ohio
State University, Columbus, OH 43210, USA}

\author{Nandini Trivedi}
\affiliation{Department of Physics, The Ohio State University,
Columbus, OH 43210, USA}

\author{Hae-Young Kee}
\affiliation{Department of Physics, University of Toronto,
Toronto, Ontario, Canada M5S 1A7} \affiliation{Canadian Institute
for Advanced Research, CIFAR Program in Quantum Materials,
Toronto, Ontario, Canada M5S 1M1}

\author{P. Chris Hammel}\email{hammel@physics.osu.edu}
\affiliation{Department of Physics, The Ohio State University,
Columbus, OH 43210, USA}

\date{\today}

\begin{abstract}
CrX$_3$ (X = Cl, Br, I) have the same crystal structure and
Hamiltonian but different ligand spin-orbit coupling (SOC)
constant $\lambda_X$, providing excellent material platform
exploring for exotic two-dimensional (2D) spin orders. Their
microscopic mechanism underlying 2D spin physics remain
unestablished, along with experimental corroboration of Kitaev
exchange interaction, central to realizing topological quantum
spin liquids. Finding direct evidence for Kitaev interaction and
determining its value has been an essential but formidable
challenge in Kitaev physics. Here we report the direct Kitaev
interaction signature in magnetic anisotropy measured by
ferromagnetic resonance (FMR) spectroscopy. We present measured
values of Heisenberg $J$, Kitaev $K$, and off-diagonal symmetric
$\Gamma$ exchange interactions in CrX$_3$ determined using FMR and
exact diagonalization. $K$ and $\Gamma$ exhibit dominant
dependencies on $\lambda_X$, indicating its central role in 2D
magnetism. Our study provides a foundation for designing 2D
magnetic materials exhibiting novel behaviors by tuning intrinsic
material parameters such as SOC.
\end{abstract}

\maketitle

Since the discovery of CrI$_3$ atomic monolayer ferromagnets
\cite{Huang2017}, two-dimensional (2D) van der Waals (vdW) magnets
have attracted much attention due to their potential for hosting
exotic 2D quantum spin physics such as bosonic topologically
protected chiral edge states
\cite{OwerreJAP2016,Chen2018,Pershoguba2018,Chen2021}, Kitaev
quantum spin liquids
\cite{Banerjee2016,Kitagawa2018,Lee2020,Xu2020} and skyrmions
\cite{Behera2019}, as well as developing 2D spintronics devices
integrated with other vdW materials such as transition metal
dichalcogenides and graphene \cite{JiangShan2018,Klein2018}. The
Kitaev interaction, which must exist in chromium trihalide family
(CrX$_3$, X = Cl, Br, I) due to its crystal symmetry
\cite{Lee2020,Cen2022,Cen2023}, is the core element in realizing
topological quantum spin liquid states. While there are
experimental reports of half-quantized thermal Hall conductance
\cite{Kasahara2018,Yokoi2021} and signatures of propagating
Majorana fermions in $\alpha$-RuCl$_3$, there are also
counter-proposals \cite{Czajka2021,Bruin2022}, leaving its
existence under debate and ambiguity as to both its estimated
value and sign. Moreover, while previously observed magnetic
behaviors associated with quantum spin liquids have been
interpreted based on the Kitaev model, there is little direct
evidence of their relevance to Kitaev interaction.

The large spin-wave gap at the Dirac point observed in CrI$_3$ by
inelastic neutron scattering (INS) \cite{Chen2018} has been
considered a possible experimental signature of the Kitaev
interaction \cite{Lee2020}, but this is controversial as the
next-nearest-neighbor (NNN) Dzyaloshinskii-Moriya (DM)
\cite{Chen2018,Chen2020} interaction could also open this gap. In
any case, the Kitaev and NNN DM interactions require unreasonably
large values to account for the large Dirac gap. It has
furthermore been pointed out that the measured gap size is very
sensitive to INS experimental conditions such as sample mosaic,
resolution, and momentum integration range leading to the
possibility that gap size can be significantly overestimated,
especially at the Dirac point where spin waves rapidly disperse
\cite{Do2022,Chen2021}. Indeed, two independent INS studies
recently reported conflicting values for the size of the Dirac gap
in CrBr$_3$: 3.5 meV \cite{Cai2021} in one case versus no gap in
another \cite{Nikitin2022}. The Dirac gap size of CrI$_3$ was
adjusted from 5 meV to 2.8 meV through improved INS measurements
\cite{Chen2021}. This situation calls for a complementary approach
to obtaining a reliable value of the Kitaev interaction.

\begin{figure}
\includegraphics[width=1.0\columnwidth]{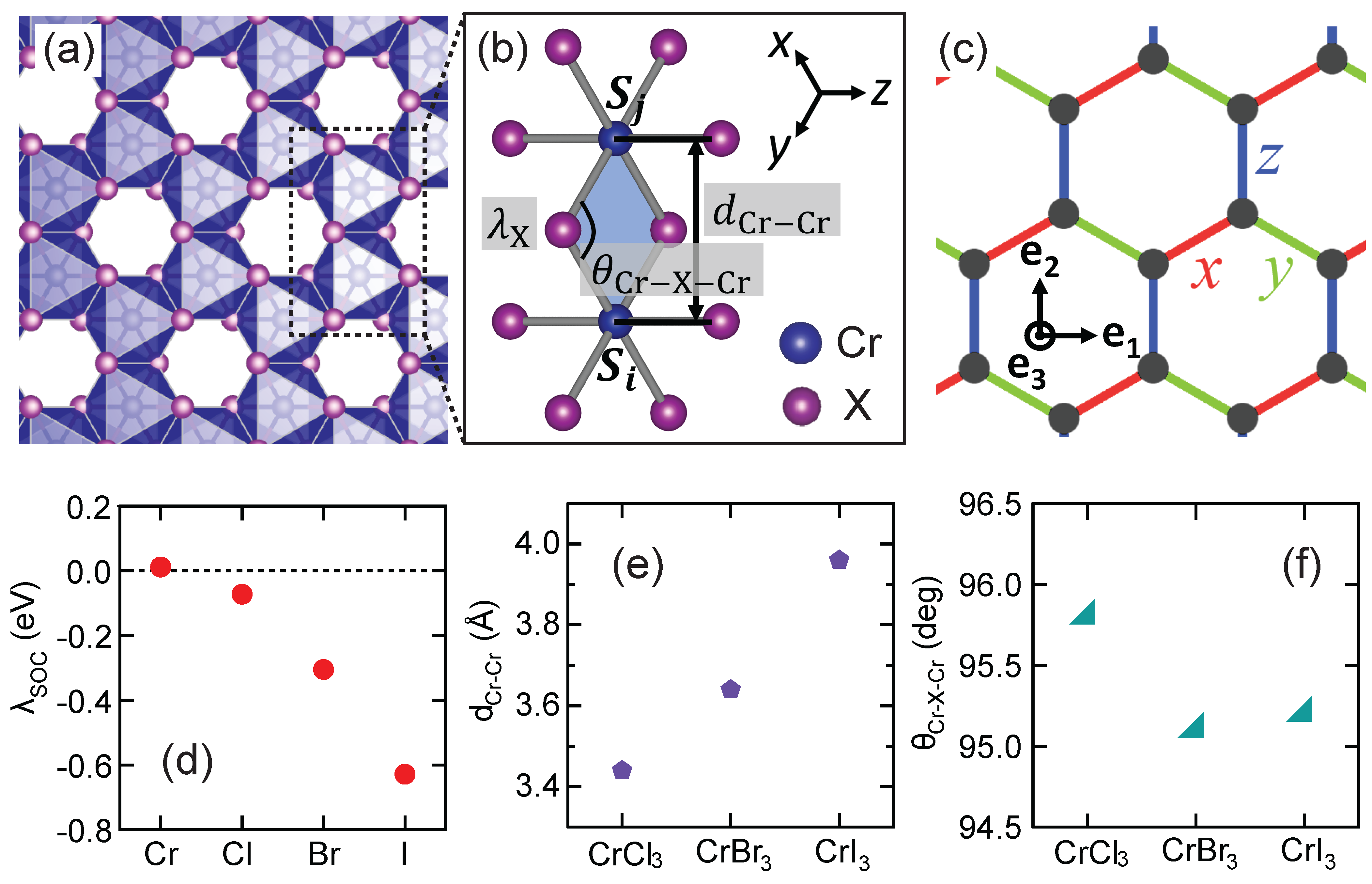}\\
\caption{(a) 2D crystal structure of CrX$_{3}$ monolayer. (b) The
parameters regarding the superexchange interaction via X$^{-}$
ligand ions between two Cr$^{3+}$ ion spins of $\mathbf{S_{i}}$
and $\mathbf{S_{j}}$ in the neighboring octahedra along z-bond:
ligand SOC constant $\lambda_{\rm{X}}$, Cr-X-Cr bond angle
$\theta_{\rm{Cr-X-Cr}}$, and Cr-Cr distance $d_{\rm{Cr-Cr}}$. (c)
2D honeycomb lattice spin model having $x$-, $y$-, and $z$-bond
dependent spin interactions. (d) The spin-orbit coupling constant
$\lambda_{\rm{SOC}}$ of the atomic orbital obtained from atomic
optical spectroscopy \cite{Moore1971,Lado2017,Kim2019}. (e) Cr--Cr
distance $d_{\rm{Cr-Cr}}$ \cite{McGuire2017}. (f) Cr--X--Cr bond
angle $\theta_{\rm{Cr-X-Cr}}$ \cite{Webster2018}. }
\label{fig:spinsystem}
\end{figure}

Ferromagnetic resonance (FMR) is a high resolution ($\sim \mu$eV)
spectroscopic tool that enables determination of the Kitaev
interaction through accurate measurement of the interaction of the
collection of ordered spins with both internal and external
environments. Similar magnetic resonance techniques have been
applied to spin excitations associated with fractionalized
Majorana excitations in $\alpha$-RuCl$_{3}$
\cite{Wellm2018,Ponomaryov2020}. The first FMR study of CrI$_3$
described the global coherent spin dynamics of the sample in
magnetic resonance by applying mean field theory to the
Hamiltonian, converting the multi-spin interaction problem into a
single spin problem \cite{Lee2020}. However, this approximation
removes all the anisotropic components of the Kitaev interaction,
rendering it isotropic and thus indistinguishable from Heisenberg
exchange. Recent theoretical studies based on symmetry analysis
showed that the Kitaev interaction leads to anisotropy in the
magnetic response between $\mathbf{e_{1}}$ and $\mathbf{e_{2}}$
directions. Kitaev interaction also leads to an anisotropy of the
FMR frequency depending on the orientation of the magnetic field
within the $\mathbf{e_{1}}$--$\mathbf{e_{3}}$ plane as indicated
in Fig. \ref{fig:spinsystem}(c) \cite{Cen2022,Cen2023}.

The CrX$_3$ materials have the same crystal structure and are
described by the same Hamiltonian as A$_2$IrO$_3$ (A = Na, Li)
\cite{Singh2012,Gretarsson2013} and $\alpha$-RuCl$_3$
\cite{Banerjee2016}, known potential honeycomb Kitaev materials,
but CrX$_3$ has contrasting features. For A$_2$IrO$_3$ and
$\alpha$-RuCl$_3$, the transition metal ion has effective spin-1/2
and dominant SOC, $\lambda_{\rm{M}} \mathbf{L} \cdot \mathbf{S}$
where $\lambda_{\rm{M}}$ plays an essential role in determining
exchange interactions. On the other hand, CrX$_3$ has spin-3/2,
and $\lambda_{\rm{X}}$, the SOC of the $p$-orbital of ligand atom
X, is considered to be the main source of superexchange
interactions. In this regard, a theoretical microscopic analysis
of CrX$_3$ was recently performed to find the origin of spin
interactions and showed that indeed Kitaev interaction can arise
for 3/2-spin by the ligand SOC \cite{Stavropoulos2021}.

CrX$_3$ (X = Cl, Br, I) share a common crystal structure and a
common Hamiltonian, but have different $\lambda_{\rm{X}}$.
Therefore, $\lambda_{\rm{X}}$ is the key single parameter that can
characterize this well-defined 2D magnetic platform once its
relationship with Hamiltonian's spin interaction constants is
clearly established. Nevertheless, there has been no systematic
experimental study to elucidate this.

Here we present the measured spin interaction constants for three
CrX$_3$ compounds determined using field-angle dependent FMR
spectroscopy and exact diagonalization (ED). Furthermore, we
investigate the relationship between those values and the ligand
SOC $\lambda_{\rm{X}}$. The magnetic anisotropy distinctively
originating from the Kitaev interaction appears in FMR spectra as
unique experimental signature that is strongly dependent on
$\lambda_{\rm{X}}$. Unlike mean field theory, ED directly
incorporates bond-dependent spin-spin interactions for multiple
spins, allowing us to determine the Kitaev interaction for CrX$_3$
from these experimental data.

CrX$_3$ spin system can be described with a 2D honeycomb lattice
spin model which has bond-dependent anisotropic exchange
interactions (see Fig. \ref{fig:spinsystem}(c)). CrX$_3$ have the
edge-sharing octahedral 2D crystal structure in Fig.
\ref{fig:spinsystem}(a) and their spin model is based on the
anisotropic superexchange interactions between two Cr spins via
Cr-X-Cr bonds arising from the SOC of ligand X (see Fig.
\ref{fig:spinsystem}(b)). Based on the crystal symmetries, the
Hamiltonian is
\begin{equation}
\mathcal{H} = \mathcal{H}_\text{E} + \mathcal{H}_\text{D} +
\mathcal{H}_\text{Z}, \label{eqn:Hamiltonian}
\end{equation}
where
\begin{align}
\mathcal{H}_{\text{E}} = \sum_{\braket{ij}\in\lambda\mu(\nu)}
[J\mathbf{S}_i\cdot\mathbf{S}_j + K S_i^\nu S_j^\nu +
\Gamma(S_i^\lambda S_j^\mu + S_i^\mu S_j^\lambda)]
\label{eqn:HamiltonianExchange}
\end{align}
describes exchange interactions,
\begin{align}
\mathcal{H}_{\text{D}} = \sum_{i > j} \frac{g^2
\mu_\text{B}^2}{r_{ij}^3} [\mathbf{S}_i\cdot\mathbf{S}_j -
\frac{3}{r_{ij}^2} \left( \mathbf{S}_i\cdot\mathbf{r_{ij}} \right)
\left( \mathbf{S}_j\cdot\mathbf{r_{ij}} \right)]
\label{eqn:HamiltonianDipole}
\end{align}
describes dipole-dipole interactions and
\begin{align}
\mathcal{H}_{\text{Z}} = - g \mu_\text{B}
\mathbf{H}_0\cdot\sum_i\mathbf{S}_i \label{eqn:HamiltonianZeeman}
\end{align}
describes Zeeman interactions. $\mathbf{S_{i}}$ is the spin-3/2
operator for the Cr$^{3+}$ ion at site $i$.
$\braket{ij}\in\lambda\mu(\nu)$ denotes that the Cr$^{3+}$ ions at
the neighboring sites $i,j$ are interacting via a $\nu$-bond,
where $\lambda,\mu,\nu\in\{x,y,z\}$. $g$ is the $g$-factor of
Cr$^{3+}$, $\mu_\text{B}$ is the Bohr magneton, and
$\mathbf{r_{ij}}$ is the distance vector joining spins at site $i$
and $j$. The magnetic anisotropy of CrX$_3$ is contributed by
$\mathcal{H}_{\rm{E}}$ and $\mathcal{H}_{\rm{D}}$ as
magnetocrystalline and shape anisotropy, respectively.

\begin{figure*}
\includegraphics[width=2.0\columnwidth]{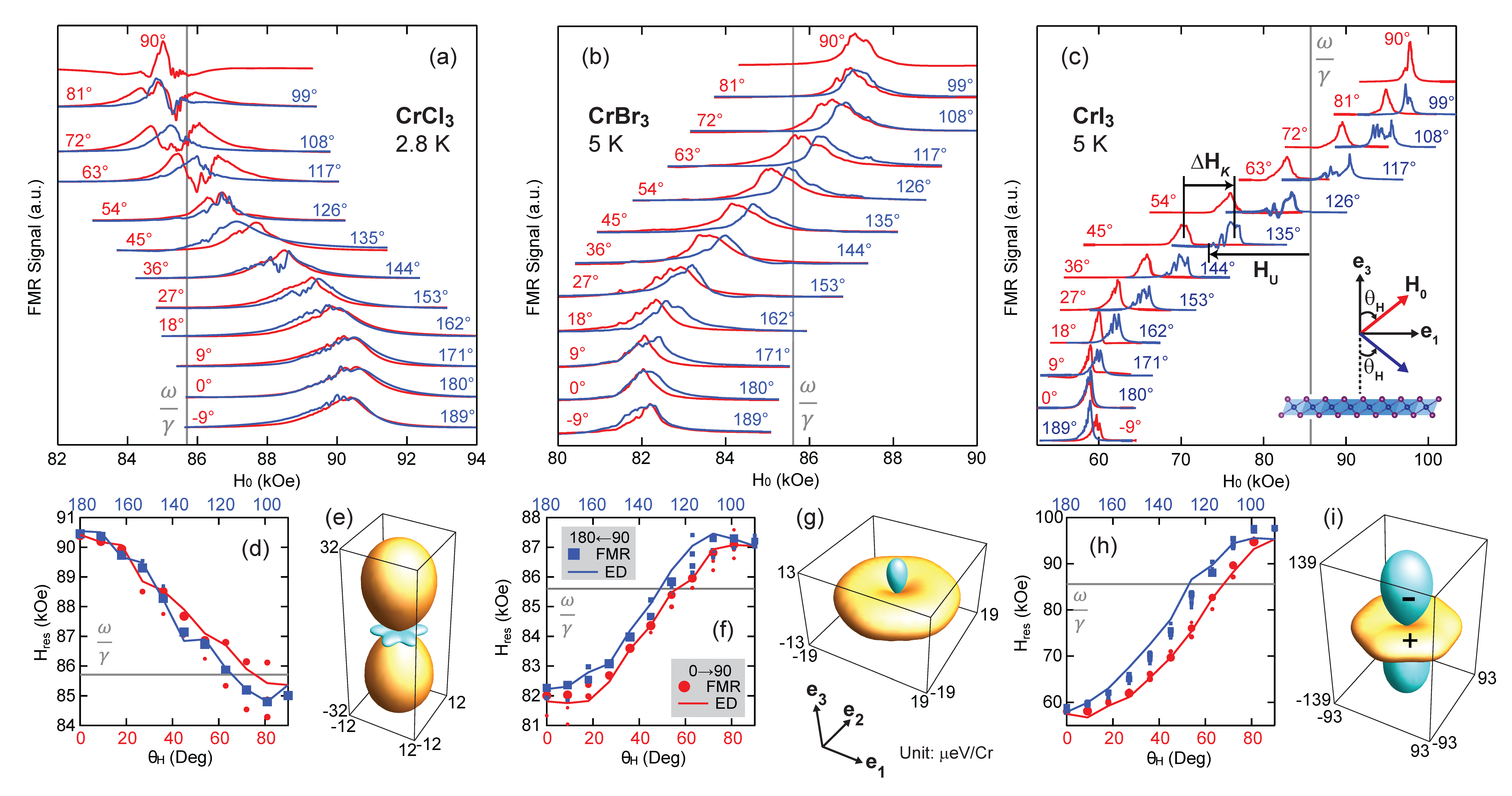}
\caption{(a)--(c) FMR spectra of CrCl$_{3}$, CrBr$_{3}$, and
CrI$_{3}$ with varying $\theta_{\rm{H}}$ describing the
orientation of magnetic field $\mathbf{H_0}$ with respect to
$\mathbf{e_{3}}$ in the $\mathbf{e_{1}}$--$\mathbf{e_{3}}$ plane
(see the inset of (c)). Each spectrum is offset and scaled
moderately for clarity. The gray line labelled $\omega /\gamma =
240$ GHz represents the resonance field $H_{\rm{res}}$
corresponding to the applied microwave frequency for the free
single ion spin. The shift of $H_{\rm{res}}$ is characterized by
two quantities: $H_{\rm{U}} \left( \theta_{\rm{H}} \right) =
H_{\rm{res}} \left( \theta_{\rm{H}} \right) - \omega / \gamma$
arising from $\Gamma$ interaction and dipole-dipole interaction in
Eq.\ (\ref{eqn:HamiltonianDipole}), and $\Delta H_{K} \left(
\theta_{\rm{H}} \right) = H_{\rm{res}} \left( 180 ^\circ -
\theta_{\rm{H}} \right) - H_{\rm{res}} \left( \theta_{\rm{H}}
\right)$ arising from the asymmetric Kitaev interaction about
$\theta_{\rm{H}}$ and $180^\circ - \theta_{\rm{H}}$, indicated by
the red and blue arrows, respectively, in the inset of (c). (d)
$H_{\rm{res}}$ vs.\ $\theta_{\rm{H}}$ extracted from (a) for
CrCl$_{3}$. (e) Total (magnetocrystalline and shape) magnetic
anisotropy energy $F \left( \theta, \phi \right)$ of CrCl$_{3}$ as
a function of spherical angles $\theta$ and $\phi$ constructed
from $H_{\rm{res}}$ in (d) using Landau theory\cite{Lee2020}. (f)
$H_{\rm{res}}$ vs. $\theta_{\rm{H}}$, and (g) $F \left( \theta,
\phi \right)$ for CrBr$_{3}$. (h) $H_{\rm{res}}$ vs.
$\theta_{\rm{H}}$, and (i) $F \left( \theta, \phi \right)$ for
CrI$_{3}$. In (d), (f), and (h), the symbol size indicates the
signal peak area in Lorentzian fits to the FMR spectra and solid
lines are exact diagonalization (ED) calculation results. In (e),
(g), and (i), orange (cyan) represents positive (negative) values.
Panel (c) is adapted from Ref. \cite{Lee2020}}.
\label{fig:FMR_data}
\end{figure*}

We determine $J$, $K$, and $\Gamma$ values from the magnetic
anisotropies of CrX$_3$ measured from the dependencies of their
FMR spectra on the magnetic field direction using a sub-THz
heterodyne quasi-optical electron spin resonance spectrometer
\cite{PRLSM2024}. FMR spectra are obtained at various
$\theta_{\rm{H}}$, the angle between the applied magnetic field
$\mathbf{H_{0}}$ and $\mathbf{e_{3}}$, in the
$\mathbf{e_{1}}$--$\mathbf{e_{3}}$ plane (see Fig.
\ref{fig:FMR_data}(c) inset). The applied electromagnetic
excitation frequency is $\omega / 2 \pi =$240 GHz. Figs.
\ref{fig:FMR_data}(a)--\ref{fig:FMR_data}(c) show the evolution of
the FMR signal of CrX$_3$ as a function of $H_0$ for a series of
orientations $\left(\theta_{\rm{H}} \right)$. We obtain the
resonance field $H_{\rm{res}}$ from Lorentzian fits to these
spectra. The evolution of $H_{\rm{res}} \left(
\theta_{\rm{H}}\right)$ for CrX$_3$ is presented in Fig.
\ref{fig:FMR_data}(d), \ref{fig:FMR_data}(f), and
\ref{fig:FMR_data}(h). The salient features of this anisotropic
behavior is best seen by considering two quantities: $H_{\rm{U}}
\left( \theta_{\rm{H}}\right) = H_{\rm{res}} \left(
\theta_{\rm{H}} \right) - \omega / \gamma$ and $\Delta H_{K}
\left( \theta_{\rm{H}} \right) = H_{\rm{res}} \left( 180 ^\circ -
\theta_{\rm{H}} \right) - H_{\rm{res}} \left( \theta_{\rm{H}}
\right)$ (see \ref{fig:FMR_data}(c)).

$H_{\rm{U}}$ reveals the uniaxial magnetic anisotropy along
$\mathbf{e_{3}}$ arising from the combination of the $\Gamma$
interaction in Eq.\ (\ref{eqn:HamiltonianExchange}) and the
dipole-dipole interaction (shape anisotropy) in Eq.\
(\ref{eqn:HamiltonianDipole}). FMR directly measures the magnitude
and polarity of the uniaxial magnetic anisotropy given by $\Delta
H_{\rm{U}} = H_{\rm{res}} \left( 90^\circ \right) - H_{\rm{res}}
\left( 0^\circ \right)$:  $-$5 kOe for CrCl$_3$, $+$5 kOe for
CrBr$_3$, and $+$35 kOe for CrI$_3$. $\Delta H_{\rm{U}}$ is
negative for CrCl$_{3}$ but positive for CrBr$_{3}$ and CrI$_{3}$
indicating their opposite polarities.

$\Delta H_{K}$ is due to the non-uniaxial magnetic anisotropy
arising from the asymmetric Kitaev interaction about two symmetric
angles $\theta_{\rm{H}}$ and $180^\circ - \theta_{\rm{H}}$,
indicated by the red and blue arrows, respectively, in the inset
of Fig. \ref{fig:FMR_data}(c). The sign of $\Delta H_{K}$ inverts
at $\theta_{\rm{H}} = 0^\circ$ and 180$^\circ$, such that
$H_{\rm{res}} \left( 9^{\circ} \right)$ is lower than
$H_{\rm{res}} \left( 171^{\circ} \right)$, but $H_{\rm{res}}
\left( -9^{\circ} \right)$ is higher than $H_{\rm{res}} \left(
189^{\circ} \right)$ (see Fig. \ref{fig:FMR_data}(b) and
\ref{fig:FMR_data}(c)). This is consistent with the $\pi$-rotation
symmetry for $\mathbf{e_{2}}$ of the anisotropic Kitaev
interaction ($H_{\rm{res}} \left( \theta_{\rm{H}} \right) =
H_{\rm{res}} \left( \theta_{\rm{H}} + 180^{\circ} \right) \neq
H_{\rm{res}} \left( \theta_{\rm{H}} - 180^{\circ} \right)$)
\cite{Cen2022}. The magnitude of $\Delta H_{K}$ is minimum for
CrCl$_{3}$, increases in CrBr$_{3}$, and maximum in CrI$_{3}$,
corresponding to the increasing strength of $\lambda_{\rm{X}}$.
This is the direct observation of the Kitaev interaction signature
in CrX$_{3}$ through magnetic anisotropy.

We describe the magnetic anisotropy of CrX$_3$ measured from FMR
as $F \left( \theta, \phi \right)$ for two spherical angles
$\theta$ and $\phi$ (see Fig. \ref{fig:FMR_data}(e) for
CrCl$_{3}$, \ref{fig:FMR_data}(g) for CrBr$_{3}$, and
\ref{fig:FMR_data}(i) for CrI$_{3}$). These are constructed from
$H_{\rm{res}} \left( \theta_{\rm{H}} \right)$ using Landau theory,
where $F \left( \theta, \phi \right)$ is expressed in terms of
classical magnetization using mean field theory and used to
calculate the resonant frequency (see Section VII of the
Supplemental Material\cite{Lee2020}). The uniaxial magnetic
anisotropy due to magnetocrystalline and shape anisotropy
dominates in CrX$_3$. The $\mathbf{e_{3}}$ (out-of-plane) axis is
the easy axis for CrBr$_3$ and CrI$_3$, but the hard axis for
CrCl$_3$.

\begin{figure}
\includegraphics[width=1.0\columnwidth]{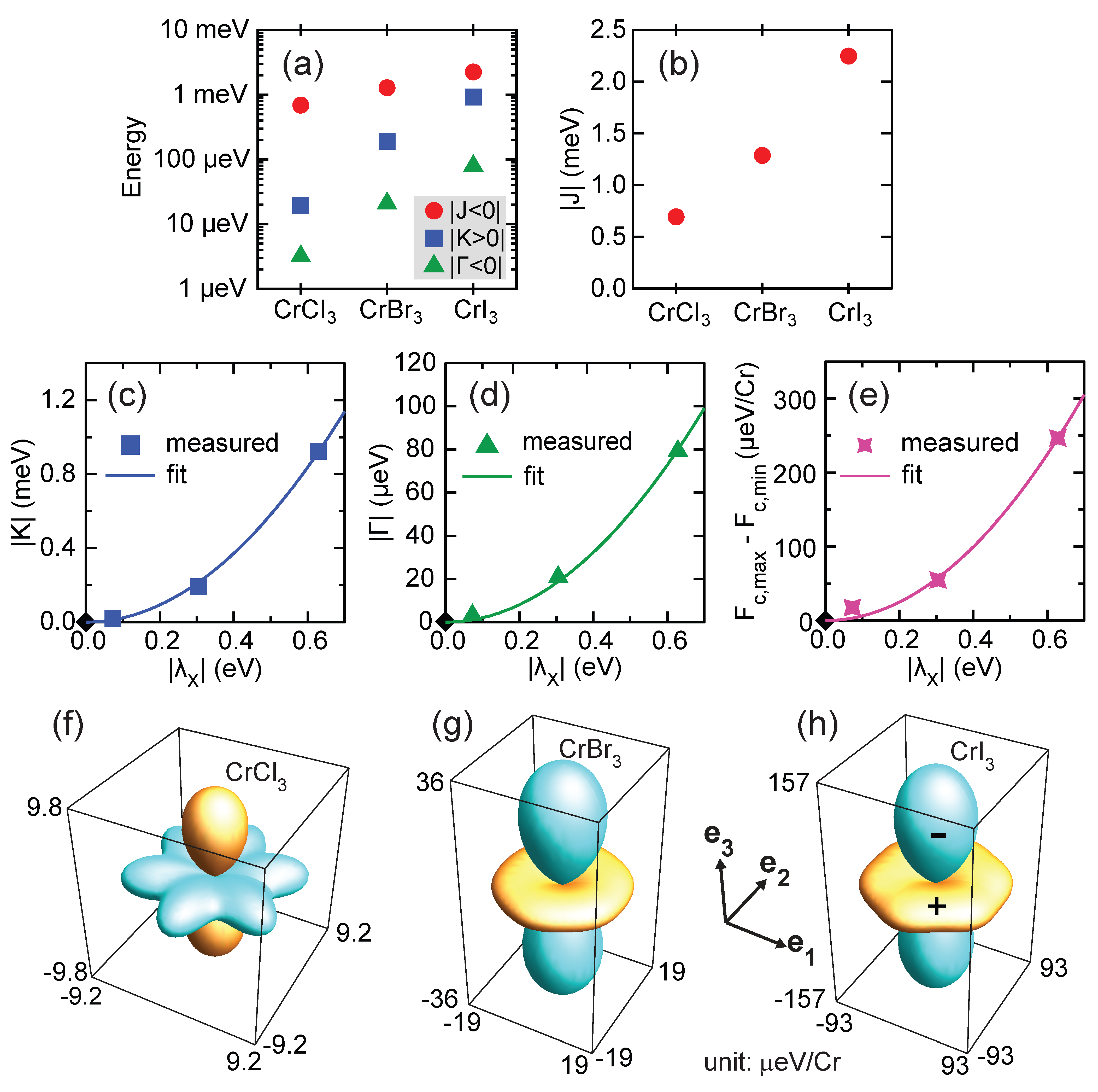}
\caption{(a) Comparison of absolute values of Heisenberg $J$,
Kitaev $K$, and off-diagonal symmetric $\Gamma$ exchange
interactions for CrX$_3$. (b) $\left| J \right|$ values for
CrX$_3$. (c) $|K|$, (d) $|\Gamma |$, and (e) $F_{\rm{c, max}} -
F_{\rm{c, min}}$ values for three $|\lambda_{\rm{X}}|$ values
corresponding to CrX$_3$ in Fig. \ref{fig:spinsystem}(d) and their
quadratic monomial fits (solid lines) based on the theory of
\cite{Stavropoulos2021} with one more data point (black diamond)
added, assuming that $K$, $\Gamma$, and $F_{\rm{c, max}} -
F_{\rm{c, min}}$ become zero when $\lambda_{\rm{X}} = 0$. Here
$F_{\rm{max}}$ and $F_{\rm{min}}$ are the maximum and minimum
values of $F_{\rm{c}} \left( \theta , \phi \right)$ for each
CrX$_3$ in (f)--(h). (f)--(h) Magnetocrystalline anisotropy energy
$F_{\rm{c}}\left( \theta , \phi \right)$ of CrCl$_3$, CrBr$_3$,
and CrI$_3$ obtained after subtracting the shape anisotropy from
$F$ in Fig. \ref{fig:FMR_data}(e), \ref{fig:FMR_data}(g), and
\ref{fig:FMR_data}(i), respectively. Orange (cyan) represents
positive (negative) values.} \label{fig:spinparameters}
\end{figure}

\begin{table}[b]
\caption{\label{tab:interaction values} Spin exchange interaction
constant values in the Hamiltonian
Eq.~(\ref{eqn:HamiltonianExchange}) of CrX$_3$ determined using
FMR and ED}
\begin{ruledtabular}
\begin{tabular}{c|ccc}
Coupling& &Value (meV)& \\
 Constant&CrCl$_3$&CrBr$_3$& CrI$_3$ \\
  \hline\\[-2mm]
  $J$ & -0.69 & -1.3 & -2.2 \\
  $K$ & 0.019 & 0.19 & 0.92 \\
  $\Gamma$ & -0.0032 & -0.021 & -0.079\\
\end{tabular}
\end{ruledtabular}
\end{table}

We determine the values of $J$, $K$, and $\Gamma$ of the
Hamiltonian in Eq.\ (\ref{eqn:HamiltonianExchange}) by fitting our
FMR data $H_{\rm{res}} (\theta_{\rm{H}})$ to the values obtained
from ED calculations with 12 sites of $S = 3/2$ for 240 GHz
\cite{PRLSM2024}. ED provides two sets of values for $J$, $K$, and
$\Gamma$ corresponding to $K>0$ and $K<0$ for CrX$_3$. According
to a recent microscopic theory for CrX$_3$
\cite{Stavropoulos2021}, $J$ and $K$ have opposite signs, with
$(J>0,K<0)$ corresponding to $t_{2g}$-$t_{2g}$ interaction and
$(J<0,K>0)$ to $e_{g}$--$t_{2g}$ interaction. Since $J$ is
negative for both sets we obtain, $K$ must be positive, and
$e_{g}$--$t_{2g}$ interaction is thought to dominate in these
ferromagnetic systems. This further implies that the superexchange
processes via $p$-orbitals of ligand X play a crucial role in
determining spin interactions.

The values of $J$, $K$, and $\Gamma$ for $K > 0$ in Figs.
\ref{fig:spinparameters}(a)--\ref{fig:spinparameters}(d) and Table
\ref{tab:interaction values} generate values that well match the
FMR data for $H_{\rm{res}} ( \theta_{\rm{H}} )$ shown in Fig.
\ref{fig:FMR_data}(d), \ref{fig:FMR_data}(f), and
\ref{fig:FMR_data}(h). Fig. \ref{fig:spinparameters}(a) shows the
relative energy scale of these values:  $|J|\gg|K| \gg |\Gamma |$
for all three CrX$_3$ compounds. This is consistent with the
energy scale of $|J| \gg |K| ( \sim r^2 |J| ) \gg |\Gamma | ( \sim
0)$ that emerges from recent microscopic second-order perturbation
theory for $S = 3/2$, where $r = \lambda_{\rm{X}} / \Delta_{pd}$
and $\Delta_{pd}$ is the atomic energy difference between the
transition metal and ligand sites \cite{Stavropoulos2021}. The $K$
and $\Gamma$ values for CrI$_3$ here are in reasonable agreement
(within 20 \%) with these exchange constants calculated using
density functional theory (DFT)\cite{Bandyopadhyay2022}.

Next, we determine the universal dependencies of $K$ and $\Gamma$
in Fig. \ref{fig:spinparameters} on the absolute value of the
ligand SOC, $|\lambda_{\rm{X}}|$. As shown in Fig.
\ref{fig:spinsystem}(d), $|\lambda_{\rm{X}}|$ increases in order
of increasing ligand mass: Cl, Br, and I. And all are much larger
than the Cr 3$d$-orbital $|\lambda_{\rm{M}}|$ which is ignored in
our analysis.

A key result is the observation that both $K$ and $\Gamma$ depend
exclusively on a single parameter: $\lambda_{\rm{X}}$. This
highlights the central role $\lambda_{\rm{X}}$ plays in
superexchange interactions. We describe the $\lambda_{\rm{X}}$
dependence of $K$ and $\Gamma$ in Fig. \ref{fig:spinparameters}(c)
and \ref{fig:spinparameters}(d) using the quadratic relationship
revealed in the recent microscopic second-order perturbation
theory \cite{Stavropoulos2021}. Figs.
\ref{fig:spinparameters}(f)--\ref{fig:spinparameters}(h) show the
magnetocrystalline anisotropy energy $F_{\rm{c}} \left( \theta ,
\phi \right)$ for CrX$_3$ as indicated, where shape anisotropy is
excluded. Fig. \ref{fig:spinparameters}(e) shows $F_{\rm{c, max}}
- F_{\rm{c, min}}$ as a function of $| \lambda_{\rm{X}} |$, which
mainly reflects the size of uniaxial magnetocrystalline anisotropy
energy, where $F_{\rm{c, max}}$ and $F_{\rm{c, min}}$ are the
maxima and minima of $F_{ \rm{c}} (\theta , \phi)$. This shows the
direct, experimentally obtained relationship between macroscopic
magnetic anisotropy and microscopic SOC $\lambda_{\rm{X}}$ arising
from the ligand atomic $p$-orbital, which is also well described
by a simple quadratic monomial, as is the case for $K$ and
$\Gamma$.

Crystal structure parameters such as Cr--Cr distance
$d_{\rm{Cr-Cr}}$ in Fig. \ref{fig:spinsystem}(e) or Cr--X--Cr bond
angle $\theta_{\rm{Cr-X-Cr}}$ in Fig. \ref{fig:spinsystem}(f) can
have a critical effect on $K$ and $\Gamma$ values as well as $J$,
but none of them shows any noticeable correlation. Recent DFT
calculations show that deformation of the monolayer crystal
structures of CrX$_3$ can sensitively influence the magnitude and
sign of the exchange interaction parameters
\cite{Webster2018,Wu2019,Pizzochero2020}. Indeed, changes in
magnetic and electronic properties due to pressure-induced crystal
deformation have been observed, such as increases in $T_{\rm{C}}$
\cite{Mondal2019}, anomalous magnetoresistance \cite{Ghosh2022},
and semiconductor-to-metal transition \cite{Ghosh2022}. In
general, superexchange interaction is known to decrease very
sensitively, changing by an order of magnitude with sub-{\AA{}}
increases in the spin-spin separation, typically exhibiting an
exponential or inverse power-law dependence
\cite{Coffman1979,Hoffmann1994}. However, this is not the case for
CrX$_3$, which exhibits the opposite behavior where the magnitudes
of $J$, $K$ and $\Gamma$ increase with increasing $d_{\rm{Cr-Cr}}$
(See Fig. \ref{fig:spinsystem}(e) and
\ref{fig:spinparameters}(a)). Also, according to the
Goodenough-Kanamori-Anderson rules
\cite{GOODENOUGH1958,KANAMORI1959,Anderson1959}, the superexchange
interaction is primarily ferromagnetic when
$\theta_{\rm{Cr-X-Cr}}$ is 90$^\circ$. In this regard, recent DFT
calculations for CrX$_3$ monolayer show that exchange interactions
are highly sensitive to small changes in $\theta_{\rm{Cr-X-Cr}}$,
and that the magnetic phase can change from ferromagnetic to
antiferromagnetic \cite{Pizzochero2020}. However,
$\theta_{\rm{Cr-X-Cr}}$ has no significant X-dependent change
around 95.5$^\circ$ and does not show a clear correlation with the
spin interaction constant values (see Fig.
\ref{fig:spinsystem}(f)), so its impact appears to be minimal.

\begin{figure}
\includegraphics[width=1.0\columnwidth]{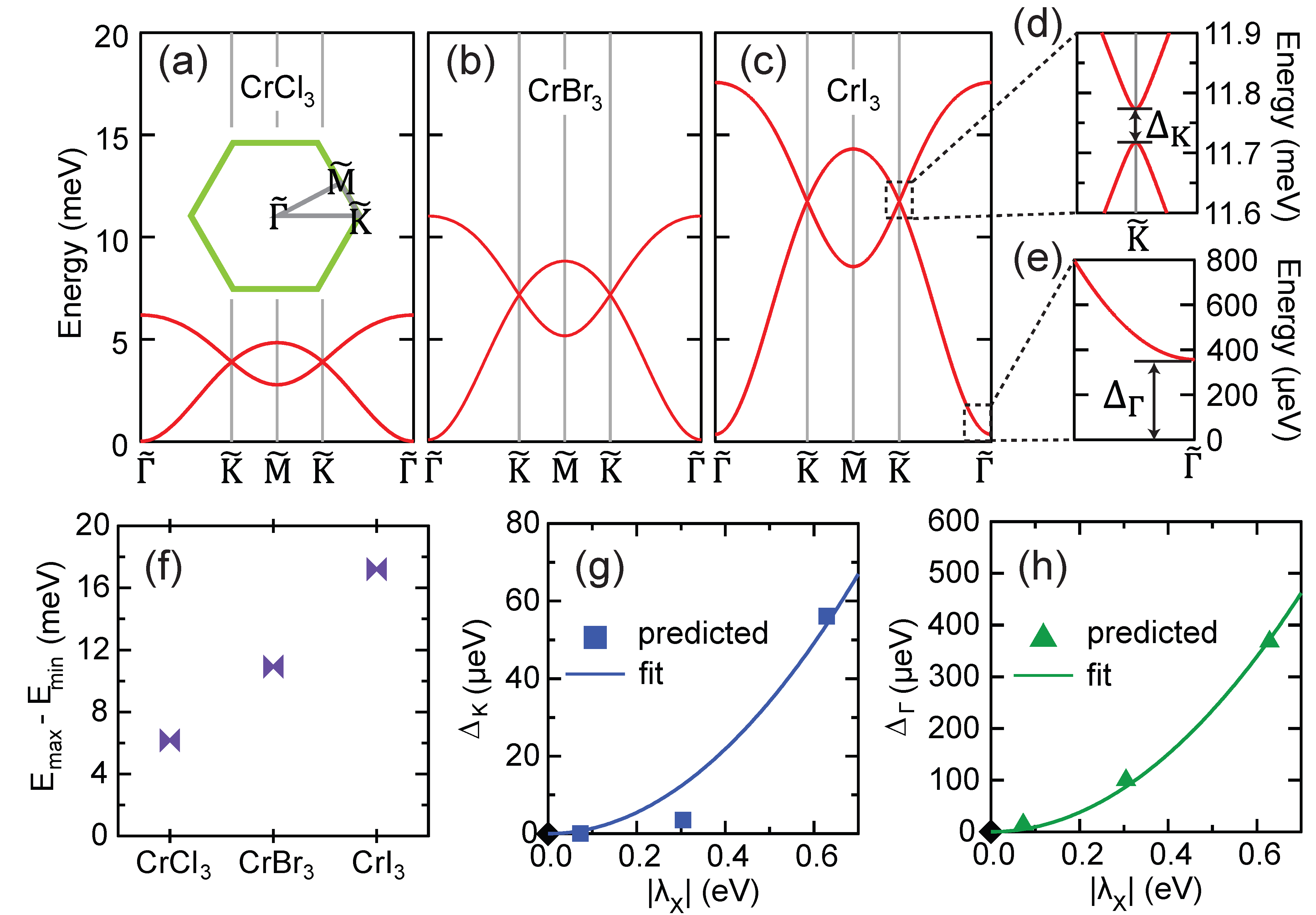}
\caption{(a)--(c) Spin-wave dispersions for CrCl$_3$, CrBr$_3$,
and CrI$_3$ predicted from linear spin-wave theory calculations
using the measured $J$, $K$, and $\Gamma$ in Fig.
\ref{fig:spinparameters} and $J_2$, NNN Heisenberg interaction.
(d) Zoom-in showing the Dirac gap $\Delta_{K}$ at $\tilde{K}$. (e)
Zoom-in on the region showing the gap $\Delta_{\Gamma}$ at
$\tilde{\Gamma}$. (f) $E_{\rm{max}} - E_{\rm{min}}$ for CrX$_3$,
where $E_{\rm{max}}$ and $E_{\rm{min}}$ are the maximum and
minimum energies of spin-wave bands in (a)--(c). (g) $\Delta_{K}$
and (h) $\Delta_{\Gamma}$ values obtained from (a)--(c) for three
$|\lambda_{\rm{X}}|$ values corresponding to CrX$_3$ in Fig.
\ref{fig:spinsystem}(d) and their quadratic monomial fits (solid
lines) with one more data point (black diamond) added, assuming
that $\Delta_{K}$ and $\Delta_{\Gamma}$ become zero when
$\lambda_{\rm{X}} = 0$. Note that DM interaction not considered
here can lead to larger $\Delta_{K}$.}
\label{fig:spinwavedispersion}
\end{figure}

Figs.
\ref{fig:spinwavedispersion}(a)--\ref{fig:spinwavedispersion}(c)
show the spin-wave dispersions for CrX$_3$ calculated using linear
spin-wave theory incorporating $J$, $K$, $\Gamma$ in Fig.
\ref{fig:spinparameters}, and $J_2$, the NNN Heisenberg
interaction. These well describe two magnon bands observed in INS,
with the coincident magnon band width $E_{\rm{max}} -
E_{\rm{min}}$, increasing from CrCl$_3$ \cite{Do2022}, through
CrBr$_3$ \cite{Nikitin2022}, to CrI$_3$ \cite{Chen2021}, where
$E_{\rm{max}}$ and $E_{\rm{max}}$ are the maximum and minimum
energies of the magnon band in Figs.
\ref{fig:spinwavedispersion}(a)--\ref{fig:spinwavedispersion}(c).
This band width is mainly determined by $|J|$ which also increases
in the order of CrCl$_3$, CrBr$_3$, and CrI$_3$ (see Fig.
\ref{fig:spinparameters}(b)).

The Kitaev interaction plays a major role in opening the Dirac gap
$\Delta_{K}$ at the momentum point $\tilde{K}$ (see Fig.
\ref{fig:spinwavedispersion}(d)). Thus, the size of $\Delta_{K}$
for $|\lambda_{\rm{X}}|$ varies roughly in accordance with a
quadratic monomial fit, as shown in Fig.
\ref{fig:spinwavedispersion}(g). The first INS reports for
CrBr$_3$ and CrI$_3$ concluded large $\Delta_{K}$ of about 3.5 meV
\cite{Cai2021} and 5 meV \cite{Chen2018}, respectively. However,
the latest INS, perhaps with reduced sample mosaic and improved
instrumental resolution, shows no Dirac gap in CrCl$_3$
\cite{Do2022} and CrBr$_3$ \cite{Nikitin2022}, which is more
consistent with the results in Fig.
\ref{fig:spinwavedispersion}(g) showing tiny $\Delta_{K}$ for all
three CrX$_3$. DM interaction could be necessary to obtain the
correct gap near Dirac point, not captured by FMR, and not
included in the spin-wave plot in Fig.
\ref{fig:spinwavedispersion}. INS suggests a 2.8 meV gap for
CrI$_3$, which is consistent with 0.2 meV DM interaction extracted
from DFT \cite{Bandyopadhyay2022}.

$\Gamma$ opens a gap $\Delta_{\Gamma}$ at the zero-momentum point
$\tilde{\Gamma}$ that overcomes Mermin-Wagner theorem by
suppressing low-energy magnon excitations thus enabling 2D
long-range ferromagnetic order. The  sizes of $\Delta_{\Gamma} =
-3 S \Gamma$ for CrX$_3$ are shown in Fig.
\ref{fig:spinwavedispersion}(h), where $\Delta_{\Gamma}$ increases
quadratically with increasing $|\lambda_{\rm{X}}|$, indicating
that it originates from the same ligand SOC as $K$ and $F_{\rm{c,
max}} - F_{\rm{c, min}}$. For CrI$_3$ we obtain $\Delta_{\Gamma} =
0.36$ meV, very close to the value, 0.37 meV, obtained recently
from high-resolution INS \cite{Chen2020}. Although not discussed
here, single-ion anisotropy $\sim S_{z}^{2}$ can also cause
uniaxial magnetic anisotropy with the same $\cos^2 \theta$ angular
dependence of energy as $\Gamma$, so its effect is
indistinguishable from $\Gamma$ in field-angle dependent FMR
experiments. However, performing ED calculations using single-ion
anisotropy results in a larger gap of 0.61 meV at
$\tilde{\Gamma}$, which is somewhat inconsistent with INS.
Therefore, it seems the effect of single-ion anisotropy is small,
and that $\Gamma$ is the primary source of the observed uniaxial
magnetocrystalline anisotropy in Figs.
\ref{fig:spinparameters}(f)--\ref{fig:spinparameters}(h). While
Heisenberg and uniaxial anisotropy $S_{z}^{2}$ terms and DM play
some roles in governing magnetism in CrI$_3$, the terms such as $K
+ \Gamma$ which break the spin conservation perpendicular to the
honeycomb plane have been shown to play a crucial role in MOKE
oscillations since any dynamics for $S_{z}$ needs breaking of spin
conservation\cite{Padmanabhan2022}.

Interestingly, the size of $\Delta_{K}$ opened by $K$ in Fig.
\ref{fig:spinwavedispersion}(g) is much smaller than the
corresponding $\left| K \right|$ value in Fig.
\ref{fig:spinparameters}(c), whereas the $\Delta_{\Gamma}$ value
in Fig. \ref{fig:spinwavedispersion}(h) is much larger than the
corresponding $\left| \Gamma \right|$ value in Fig.
\ref{fig:spinparameters}(d). This is because the large value of
$J$ significantly inhibits $K$ from opening the gap $\Delta_{K}$,
whereas $J$ does not affect the size of $\Delta_{\Gamma}$. For
example, our calculation for CrI$_3$ shows that a 50$\%$ decrease
in $\left| J \right|$, while keeping other interaction constants
the same, results in a 77$\%$ increase in $\Delta_{K}$.

In conclusion, we present measurements of the spin interaction
constants in the $J\!K\Gamma$ Hamiltonian for three CrX$_3$
compounds obtained experimentally from field angle-dependent
ferromagnetic resonance and theoretically from exact
diagonalization. We find that $K$ and $\Gamma$ depend dominantly
on the ligand SOC $\lambda_{\rm{X}}$ following the quadratic
monomial dependence predicted by microscopic second-order
perturbation theory. $J$ may suppress the effects of $K$, such as
by inhibiting the opening of the gap $\Delta_{K}$, which may make
it difficult to observe the consequences of the Kitaev
interaction, beyond the magnetic anisotropy measured by FMR and
reported here. To realize Kitaev physics, studies of $J$,
especially its physical origins and methods of suppressing it,
should be conducted in parallel with studies of $K$. Our
experimental discovery of the microscopic mechanism of 2D
magnetism in CrX$_3$ paves the way to design and explore 2D
magnetic materials exhibiting novel behaviors by tuning intrinsic
material parameters such as spin-orbit coupling.

\begin{acknowledgments}
This research was primarily supported by the Center for Emergent
Materials, an NSF MRSEC, under award number DMR-2011876. The
National High Magnetic Field Laboratory (NHMFL) is funded by the
National Science Foundation Division of Materials Research (Grants
DMR-1644779 and DMR-2128556) and the State of Florida. JC and HYK
are supported by the NSERC Discovery Grant No. 2022-04601 and the
Canada Research Chairs Program No. CRC-2019-00147. This research
was enabled in part by support provided by Compute Ontario, Calcul
Qu\'{e}bec, and the Digital Research Alliance of Canada.
\end{acknowledgments}

\bibliography{CrX3FMR}

\end{document}